\documentclass[11pt,a4paper]{article}

\usepackage[a4paper,margin=1in]{geometry}
\usepackage{setspace}
\usepackage{graphicx}

\makeatletter
\@ifundefined{pandocbounded}{%
  \newsavebox\pandoc@box
  \newcommand*\pandocbounded[1]{%
    \sbox\pandoc@box{#1}%
    \Gscale@div\@tempa{\textheight}{\dimexpr\ht\pandoc@box+\dp\pandoc@box\relax}%
    \Gscale@div\@tempb{\linewidth}{\wd\pandoc@box}%
    \ifdim\@tempb\p@<\@tempa\p@
      \let\@tempa\@tempb
    \fi
    \ifdim\@tempa\p@<\p@
      \scalebox{\@tempa}{\usebox\pandoc@box}%
    \else
      \usebox{\pandoc@box}%
    \fi
  }%
}{}
\makeatother

\usepackage{fontspec}

\usepackage{amsmath,amssymb,amsfonts,bm,mathrsfs}

\usepackage{booktabs,multirow,array,longtable}
\usepackage{caption}
\graphicspath{{./}{figures/}}

\usepackage{enumitem}
\usepackage{xurl}

\usepackage[numbers,sort&compress]{natbib}
\usepackage{xcolor}
\usepackage[hidelinks]{hyperref}

\title{\bfseries A Candidate Framework for Free-Space Quantum Key Distribution based on Geometrical-Configuration Modulation}

\author{
Yu-Ming Bai$^{1,2}$, Yu-Xuan Liu$^{1,2}$, Ming-Han Ding$^{1,2}$ and Jun-Lin Li$^{1,2,*}$\\[2mm]
\small $^{1}$Department of Physics, Tsinghua University, Beijing, China.\\
\small $^{2}$State Key Laboratory of Low Dimensional Quantum Physics, Tsinghua University, Beijing, China.\\[1mm]
\small $^{*}$Corresponding author: center@mail.tsinghua.edu.cn
}

\date{\today}

\begin{document}

\maketitle

\begin{abstract}
This paper proposes a candidate framework for free-space quantum key distribution (QKD) based on geometrical-configuration modulation (GM). In the minimal implementation considered here, Alice coherently splits a single photon emitted from one source into two spatial output modes with a tunable separation, and uses the source separation $R$ as the GM variable that defines the prepared single-photon spatial superposition state. Bob records the single-photon detection coordinate in the far field or Fourier plane, providing the correlated data used for soft-input information reconciliation. Based on this physical mechanism, we first establish an $R-x$ protocol model in which the source separation $R$ and the single-photon detection coordinate $x$ are random variables, and further propose an $R-\Delta x$ extension based on the difference variable $\Delta x$ between adjacent accepted detection events to mitigate slowly varying center drift in free-space links. The framework specifies state preparation, far-field conditional probabilities, soft-input information generation, parameter estimation, reconciliation, and asymptotic candidate key-rate formulas. A complete composable security analysis further requires derive an explicit computable upper bound on Eve's information from experimentally observed parameters, together with finite-key analysis and experimental validation under free-space conditions. The proposed candidate framework (GM-QKD) provides a modulation approach based on spatial degrees of freedom in which the source geometry serves as the modulation variable.

\end{abstract}


\vspace{4mm}

\tableofcontents
\vspace{1em}

\section{Introduction}\label{sec:introduction}

The central goal of free-space quantum key distribution (QKD) is to enable spatially separated Alice and Bob to extract a shared random key from the preparation, transmission, and measurement of quantum states. The BB84 protocol established the basic prepare-and-measure QKD framework based on nonorthogonal quantum states\cite{RN559}; Shor and Preskill provided a representative proof of BB84 security\cite{RN561}; privacy amplification provides an information-theoretic tool for extracting a secure key from partially leaked correlated random variables\cite{RN598}. In practical systems, weak coherent sources, channel loss, and finite-sample effects make security analysis more complex; the decoy-state method and finite-key security analysis provide standard tools for estimating single-photon contributions and evaluating key length with finite data, respectively\cite{RN596,RN574,RN600,RN599}.Continuous-variable QKD (CV-QKD) provides another important route in which continuous optical observables, typically field quadratures measured by homodyne or heterodyne detection, are used together with parameter estimation and reconciliation to extract secret keys\cite{RN572,RN573,RN609}.

The encoding degrees of freedom in QKD are not limited to polarization, phase, or time-bin. Spatial degrees of freedom have an intrinsic high-dimensional structure and have therefore long been regarded as an important direction for increasing the information capacity per photon and developing free-space quantum communication. The security and large-alphabet advantages of high-dimensional QKD have been systematically discussed theoretically\cite{RN601,RN602}; spatially encoded qudits, position-momentum variables, and orbital-angular-momentum modes have also been used in experimental studies of different forms of spatial-degree-of-freedom QKD\cite{RN603,RN604,RN605,RN606}. Meanwhile, the development of free-space and satellite QKD has shown that spatial transmission can support long-distance quantum communication, but loss, background light, pointing error, and atmospheric turbulence in practical links also directly affect parameter estimation and key generation\cite{RN569,RN607}.

In free-space optical communication, a natural question is whether the transmitter geometry itself can carry information, in addition to the amplitude, phase, polarization, or mode label of the optical field. Geometrical-configuration modulation (GM) is motivated by this perspective. Existing work on GM encodes information into the geometrical configuration of a multi-source emitting structure and its second-order correlations, and has demonstrated robust transmission in free-space links under strong turbulence when combined with active correlation decoding\cite{gmfso}. This result shows that the transmitter geometry is not merely a fixed structural parameter of an optical system, but can also become an information-bearing degree of freedom for tunable modulation.

Here we consider how GM can be used in single-photon quantum communication. We choose the minimal geometrical configuration: a tunable double-source system formed by coherently splitting a single photon emitted from one source into two spatial output modes. Alice changes the source separation $R_m$ between the two coherent output ports, thereby preparing different single-photon spatial superposition states; Bob records the single-photon detection coordinate in the far field or Fourier plane. Because the source separation determines the spatial frequency of the far-field interference distribution, different $R_m$ give rise to different conditional distributions in Bob's detection-coordinate statistics.

\begin{figure}
\centering
\includegraphics[width=\linewidth,keepaspectratio]{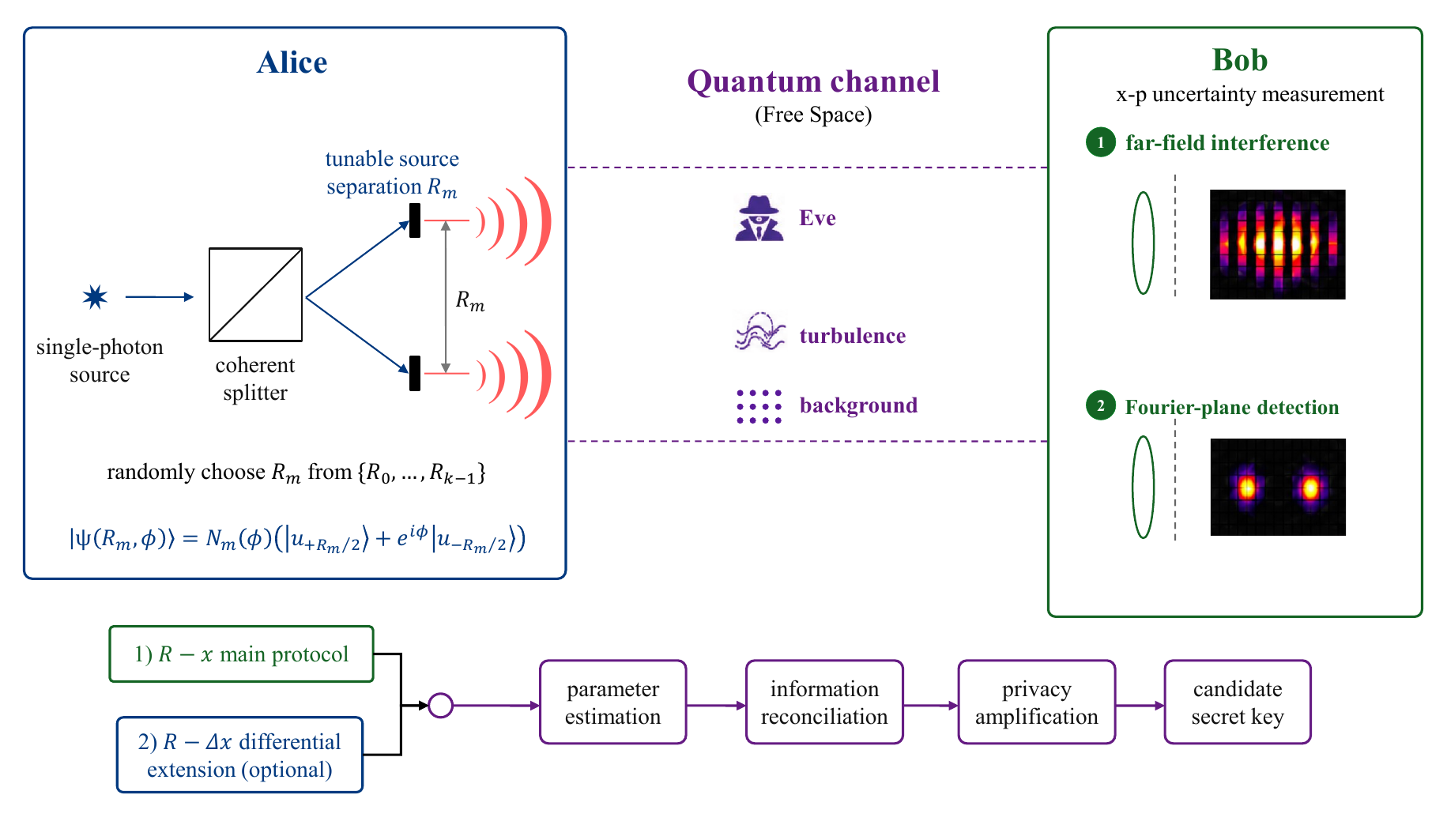}
\caption{\textbf{Schematic of the double-source single-photon interference candidate QKD framework based on geometrical-configuration modulation.}
Alice randomly selects the double-source separation $R_m$ to prepare a single-photon spatial superposition state. Bob records the single-photon detection coordinate in the far-field interference plane or Fourier plane and generates soft information about the geometrical-configuration symbol. The $R-x$ protocol uses the detection coordinate $x$ directly, whereas the $R-\Delta x$ extension uses the difference variable $\Delta x$ between adjacent accepted detection events to reduce slowly varying drift. Parameter estimation, information reconciliation, and privacy amplification form the subsequent classical post-processing steps, with security determined by the resulting bound on Eve's information.}
\label{fig:gm-qkd-schematic}
\end{figure}

Based on this physical picture, we propose two protocol variants. The first is the main $R-x$ protocol: Alice randomly selects the source separation $R_m$, and Bob directly uses the single-photon detection coordinate $x$ to generate soft information about the geometrical-configuration symbol. The second is the $R-\Delta x$ differential protocol: when a slowly varying center drift exists in the free-space link, Bob uses the difference variable $\Delta x$ between adjacent accepted detection events as the observation variable to mitigate the influence of common-mode drift on absolute detection coordinates. Existing studies on spatial-mode communication and OAM-QKD show that atmospheric turbulence can cause mode crosstalk, distribution distortion, and performance degradation\cite{RN535,RN582}. Therefore, in spatial-degree-of-freedom QKD, it is necessary to consider physical propagation perturbations, parameter estimation, and the robustness of post-processing simultaneously. Figure~\ref{fig:gm-qkd-schematic} shows the procedure of the GM-based double-source single-photon interference candidate QKD protocol.

We establish a candidate framework for QKD based on GM, provide the basic protocol procedure from state preparation and spatial detection to soft-input information reconciliation, discuss the operating regimes of the main $R-x$ protocol and the $R-\Delta x$ differential extension in free-space scenarios, and give the basic structure of parameter estimation, worst-case security optimization, and candidate key-rate formulas. The remainder of this paper is organized as follows. Section~\ref{sec:gm-single-photon-model} establishes the GM physical model of double-source single-photon interference and gives the relationship between the source separation $R_m$ and the far-field single-photon conditional probability distribution. Section~\ref{sec:rx-protocol} proposes the main $R-x$ protocol and explains geometrical-configuration modulation, spatially resolved detection, soft-input information generation, and reconciliation. Section~\ref{sec:rdeltax-protocol} proposes the $R-\Delta x$ differential protocol and discusses the construction and post-processing of the difference variable between adjacent accepted detection events. Section~\ref{sec:parameter-estimation-security} discusses parameter estimation, candidate key rates, and security considerations. Section~\ref{sec:discussion} discusses the physical role, operating regimes, and experimental considerations of the candidate protocol framework. Section~\ref{sec:conclusion} concludes the paper.

\section{Single-Photon Interference Model for Geometrical-Configuration Modulation}
\label{sec:gm-single-photon-model}

\subsection{Geometrical-Configuration Modulation Variable and State Preparation}
\label{subsec:gm-variable-state-preparation}

GM-QKD uses geometrical parameters of a coherent emitting structure as modulation degrees of freedom for quantum-state preparation. We consider its minimal implementation: a tunable double-source system obtained by coherently splitting a single photon emitted from one source into two spatial output modes. To account for finite apertures and finite beam widths, the two coherent output ports are represented below by translated transverse spatial modes.

Let $M_i$ denote Alice's geometrical-configuration symbol random variable in the $i$-th emission time slot, taking values in ${0,1,\dots,K-1}$. Its realization is denoted by $m_i$, and the corresponding realized source separation is $R_{m_i}$. Equivalently, $R_{M_i}$ denotes the random source separation before conditioning on a particular symbol value.

Let the transverse positions of the two spatial output ports be $+R_m/2$ and $-R_m/2$. In each emission time slot, Alice randomly selects the source separation $R_m$ from the public discrete alphabet
\begin{equation}
\mathcal{R}
=
\{R_0,R_1,\dots,R_{K-1}\}
\end{equation}
where $m\in\{0,1,\dots,K-1\}$ is the symbol value. Different $R_m$ correspond to different double-source single-photon spatial superposition states and give rise to different far-field conditional probability distributions for Bob.

Let the untranslated single-aperture transverse mode be $|u_0\rangle$, and denote the translated modes associated with the two output ports by $|u_{+R_m/2}\rangle$ and $|u_{-R_m/2}\rangle$. In a one-dimensional transverse model, the single-photon state prepared by Alice is written as
\begin{equation}
|\psi(R_m,\phi)\rangle
=
\mathcal{N}_m(\phi)
\left(
|u_{+R_m/2}\rangle
+
e^{i\phi}
|u_{-R_m/2}\rangle
\right),
\end{equation}
where $\phi$ is the relative phase of the double source, and the normalization factor is
\begin{equation}
\mathcal{N}_m(\phi)
=
\left[
2+
2\operatorname{Re}
\left(
e^{i\phi}
\langle u_{+R_m/2}|u_{-R_m/2}\rangle
\right)
\right]^{-1/2}.
\end{equation}
When the overlap of the two output modes is small, or when mode engineering makes them approximately orthogonal, $\mathcal{N}_m(\phi)\simeq 1/\sqrt{2}$ and the state reduces to the ideal two-path superposition. We take $R_m$ as the main geometrical-configuration modulation variable, while $\phi$ is used for phase calibration, phase checking, or subsequent extensions to joint modulation.

\subsection{Far-Field Single-Photon Probability Distribution}
\label{subsec:far-field-single-photon-probability}

Bob records the single-photon detection coordinate in the far field or Fourier plane. Let $L$ denote the propagation distance or the effective propagation scale of the Fourier system, and let $x$ denote the transverse coordinate on the detection plane. Under the far-field approximation, the relative propagation phase difference of the two spatial output ports at detection point $x$ is
\begin{equation}
\frac{kR_m x}{L},
\end{equation}
where $k=2\pi/\lambda$ is the wavenumber and $\lambda$ is the optical wavelength. Therefore, the source separation $R_m$ controls the spatial frequency of the single-photon detection-coordinate probability distribution through the far-field interference term.

We adopt the following receiver-calibrated effective far-field probability distribution as a baseline model,
\begin{equation}
P_0(x|R_m,\phi)
=
\frac{
A(x)
\left[
1+V\cos\left(\kappa_m x+\phi\right)
\right]
}{
Z(R_m,\phi)
},
\end{equation}
where
\begin{equation}
\kappa_m
=
\frac{kR_m}{L}
=
\frac{2\pi R_m}{\lambda L}.
\end{equation}
Here $A(x)$ denotes the effective spatial envelope, $V$ denotes the effective interference visibility, and $Z(R_m,\phi)$ is the normalization factor:
\begin{equation}
Z(R_m,\phi)
=
\int
A(x)
\left[
1+V\cos\left(\kappa_m x+\phi\right)
\right]
\,dx.
\end{equation}
For compactness, the possible dependence of $A(x)$ and $V$ on the geometrical configuration is not written explicitly. In experimental calibration, if changing the source separation also changes the envelope, efficiency, or visibility, the model can be generalized to $A_m(x)$ and $V_m$, or replaced directly by the calibrated discrete conditional probabilities $p_j(R_m,\phi,s)$.

\begin{figure}[t]
\centering
\includegraphics[width=0.82\linewidth]{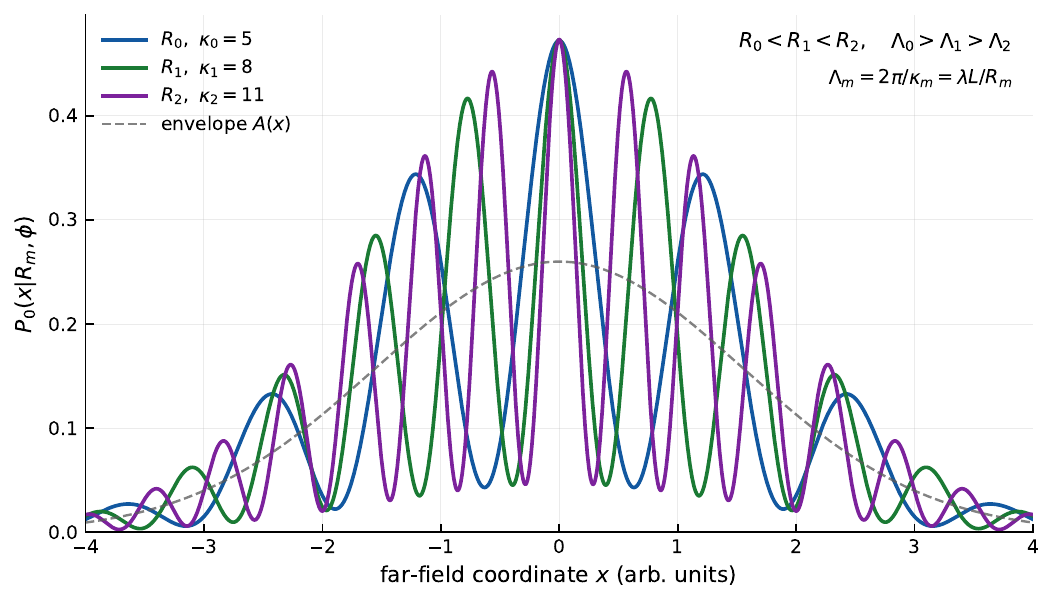}
\caption{
\textbf{Far-field single-photon conditional probability distributions corresponding to different geometrical configurations.}
The schematic uses the effective model $P_0(x|R_m,\phi)\propto A(x)[1+V\cos(\kappa_m x+\phi)]$, where $\kappa_m=2\pi R_m/(\lambda L)$. Under the same spatial envelope $A(x)$, a larger source separation $R_m$ corresponds to a larger spatial frequency $\kappa_m$ and a shorter fringe period $\Lambda_m=\lambda L/R_m$. Bob's soft information comes from the statistical distinguishability of a single-photon detection coordinate among different conditional distributions.
}
\label{fig:conditional-distributions}
\end{figure}

For fixed $\lambda$ and $L$, the corresponding fringe period is
\begin{equation}
\Lambda_m
=
\frac{2\pi}{\kappa_m}
=
\frac{\lambda L}{R_m}.
\end{equation}
Thus, Alice's discrete choice of the source separation $R_m$ appears as a discrete change of spatial frequency in Bob's single-photon detection-coordinate distribution, as shown in Fig.~\ref{fig:conditional-distributions}.

\subsection{Discrete Detection and Conditional Probability}
\label{subsec:discrete-detection-conditional-probability}

In practical free-space links, the receiver is also affected by center offset, background light, dark counts, and nonuniform detection response. Let $s$ denote the effective center offset, $B(x)$ denote the normalized spatial distribution of background events, and $\epsilon_{\mathrm{bg}}$ denote the fraction of background events. Then the observed distribution conditioned on an accepted detection event is written as
\begin{equation}
P_{\mathrm{obs}}(x|R_m,\phi,s)
=
(1-\epsilon_{\mathrm{bg}})
\frac{
A(x-s)
\left[
1+V\cos\left(\kappa_m(x-s)+\phi\right)
\right]
}{
Z(R_m,\phi,s)
}
+
\epsilon_{\mathrm{bg}}B(x),
\end{equation}
where
\begin{equation}
Z(R_m,\phi,s)
=
\int
A(x-s)
\left[
1+V\cos\left(\kappa_m(x-s)+\phi\right)
\right]
\,dx.
\end{equation}
Here $s$ characterizes slowly varying pointing error or overall center offset, $A(x)$ absorbs the single-aperture diffraction envelope, the receiving optical response, and the detector spatial response, and $V$ characterizes the effective interference visibility after propagation.

Let the $j$-th detection pixel cover the region $\mathcal{D}_j$. The discrete conditional probability used by Bob is
\begin{equation}
p_j(R_m,\phi,s)
=
P(X=x_j|R_m,\phi,s)
=
\int_{\mathcal{D}_j}
P_{\mathrm{obs}}(x|R_m,\phi,s)
\,dx.
\end{equation}
If the center offset can be estimated through tracking or calibration, the value $p_j(R_m,\phi,\hat{s})$ is used. If the center offset is only described statistically, the marginalized distribution
\begin{equation}
p_j(R_m,\phi)
=
\int
p_j(R_m,\phi,s)\rho(s)\,ds,
\end{equation}
is used, where $\rho(s)$ denotes the statistical distribution of the center offset.

\subsection{Differential Observation Variable}
\label{subsec:differential-observation-variable}

Pointing error, turbulence-induced tilt, and system drift in a free-space link change the detection coordinate at the receiver. The $i$-th accepted detection event can be written as
\begin{equation}
x_i
=
y_i+s_i+\nu_i,
\end{equation}
where $y_i$ is the ideal detection-coordinate random variable determined by the geometrical configuration $R_{m_i}$, $s_i$ denotes slowly varying center offset or common-mode drift, and $\nu_i$ denotes readout error, local perturbation, and background noise.

When $s_i$ varies slowly between adjacent accepted detection events,
\begin{equation}
s_{i+1}-s_i \approx 0.
\end{equation}
Define the differential variable
\begin{equation}
\Delta x_i
=
x_{i+1}-x_i,
\end{equation}
then
\begin{equation}
\Delta x_i
=
(y_{i+1}-y_i)
+
(s_{i+1}-s_i)
+
(\nu_{i+1}-\nu_i).
\end{equation}
Therefore, the differential variable can mitigate the influence of slowly varying common-mode drift on the observation while preserving statistical information about adjacent geometrical-configuration events.

\section{\texorpdfstring{Main $R-x$ Protocol}{Main R-x Protocol}}
\label{sec:rx-protocol}

\subsection{Protocol Structure}
\label{subsec:rx-structure}

The $R-x$ protocol is the main protocol of GM-QKD. In each emission time slot, Alice randomly selects a symbol value $m_i$ and thereby determines the double-source separation $R_{m_i}$; Bob records the single-photon detection coordinate $x_i$ or the discrete pixel index $j_i$ in the far field or Fourier plane. Since a single detection event usually carries only partial information about $R_{m_i}$, Bob generates soft information only according to the calibrated conditional probabilities and recovers Alice's geometrical-configuration symbol sequence through information reconciliation.

The basic statistical chain of this protocol is
\begin{equation}
M_i
\longrightarrow
R_{M_i}
\longrightarrow
|\psi(R_{M_i},\phi_i)\rangle
\longrightarrow
X_i ,
\end{equation}
where $M_i$ is Alice's symbol random variable and $X_i$ is Bob's discrete detection output. The corresponding single-event conditional probability is
\begin{equation}
P(X_i=x_j|M_i=m)
=
p_j(R_m,\phi_i,s_i),
\end{equation}
where $s_i$ denotes the receiver center offset. If the center offset can be estimated through tracking or calibration as $\hat{s}_i$, Bob uses $p_j(R_m,\phi_i,\hat{s}_i)$; if only its statistical distribution is known, Bob uses the marginalized effective conditional probability $p_j(R_m,\phi_i)$.

\subsection{Public Settings}
\label{subsec:rx-public-setting}

Before the protocol starts, Alice and Bob publicly agree on the geometrical-configuration alphabet
\begin{equation}
\mathcal{R}
=
\{R_0,R_1,\dots,R_{K-1}\},
\end{equation}
the symbol prior distribution
\begin{equation}
\pi_m=P(M=m),
\end{equation}
the phase calibration or phase modulation rule $\phi_i$, the set of discrete detector outputs
\begin{equation}
\mathcal{X}
=
\{x_1,x_2,\dots,x_J\},
\end{equation}
and the calibrated conditional probability model
\begin{equation}
p_j(R_m,\phi,s)
=
P(X=x_j|R_m,\phi,s).
\end{equation}
The error-correction code, the hash family for privacy amplification, the parameter-estimation rules, and the authenticated classical channel are also determined at this stage and are used for subsequent soft-input information generation, parameter estimation, and classical post-processing. If the protocol includes a near-field or other conjugate-space check path, its sampling probability, measurement setting, and announcement rules are also fixed at this stage; this check path is used only as an input for parameter estimation and does not enter raw key generation.

\subsection{Quantum Transmission and Accepted Detection Events}
\label{subsec:rx-transmission-detection}

Alice prepares the double-source single-photon state $|\psi(R_{m_i},\phi_i)\rangle$ defined in Section~\ref{subsec:gm-variable-state-preparation} according to $R_{m_i}$ and the relative phase $\phi_i$, and sends it to Bob through the free-space quantum channel.

Bob records events in the far field or Fourier plane using a spatially resolved single-photon detector array. After completing one round of transmission, Bob publicly announces the time indices of accepted detection events through the authenticated classical channel, but does not disclose the specific detection coordinates. Alice retains the corresponding transmitted symbols according to the accepted detection time slots, obtaining the accepted sample set
\begin{equation}
\mathcal{I}_{\mathrm{det}}
=
\{i:\text{the }i\text{-th time slot produces an accepted detection event}\}.
\end{equation}
Undetected time slots are treated as erasure events and do not enter the raw symbol sequence; their statistical frequency enters subsequent parameter estimation.

\subsection{Soft-Input Information Generation and Reconciliation}
\label{subsec:rx-soft-information}

For each accepted detection event $i\in\mathcal{I}_{\mathrm{det}}$, Bob records the triggered pixel $j_i$ and computes the log-likelihood of each geometrical-configuration symbol. If the center offset estimate is $\hat{s}_i$, then
\begin{equation}
L_i(m)
=
\log p_{j_i}(R_m,\phi_i,\hat{s}_i).
\end{equation}
If the marginalized model is used, then
\begin{equation}
L_i(m)
=
\log p_{j_i}(R_m,\phi_i).
\end{equation}
After taking the prior distribution into account, Bob uses
\begin{equation}
\Gamma_i(m)
=
\log \pi_m+\log p_{j_i}(R_m,\phi_i)
\end{equation}
as soft information. The corresponding posterior probability is
\begin{equation}
P(M_i=m|X_i=x_{j_i})
=
\frac{
\pi_m p_{j_i}(R_m,\phi_i)
}{
\sum_{n=0}^{K-1}
\pi_n p_{j_i}(R_n,\phi_i)
}.
\end{equation}

Bob uses
\begin{equation}
\boldsymbol{\Gamma}_i = (\Gamma_i(0),\Gamma_i(1),\dots,\Gamma_i(K-1))
\end{equation}
as the soft input for the $i$-th accepted detection event. Over all accepted detection events, these soft inputs form the sequence
\begin{equation}
\boldsymbol{\Gamma}_{1},\boldsymbol{\Gamma}_{2},\dots,\boldsymbol{\Gamma}_{n}.
\end{equation}
Alice holds the corresponding raw geometrical-configuration sequence
\begin{equation}
M_{\tau_1},M_{\tau_2},\dots,M_{\tau_n},
\end{equation}
where $\tau_i$ denotes the photon-emission time slot corresponding to the $i$-th accepted detection event.

The two parties then perform high-dimensional information reconciliation through the authenticated classical channel, so that Bob recovers Alice's accepted symbol sequence. The public information disclosed during error correction and its leakage are included in the subsequent unified security analysis.

\subsection{Protocol Features}
\label{subsec:rx-features}

The $R-x$ protocol has a direct single-event conditional probability model and is suitable as the basic protocol of GM-QKD. Its main limitation is its sensitivity to center offset and common-mode drift, so it usually needs to be combined with beam tracking, center estimation, or a differential-observation extension in free-space links.

\section{\texorpdfstring{$R-\Delta x$ Differential Protocol}{R-Delta x Differential Protocol}}
\label{sec:rdeltax-protocol}

\subsection{Protocol Structure}
\label{subsec:rdeltax-structure}

\begin{figure}[t]
\centering
\includegraphics[width=\linewidth,keepaspectratio]{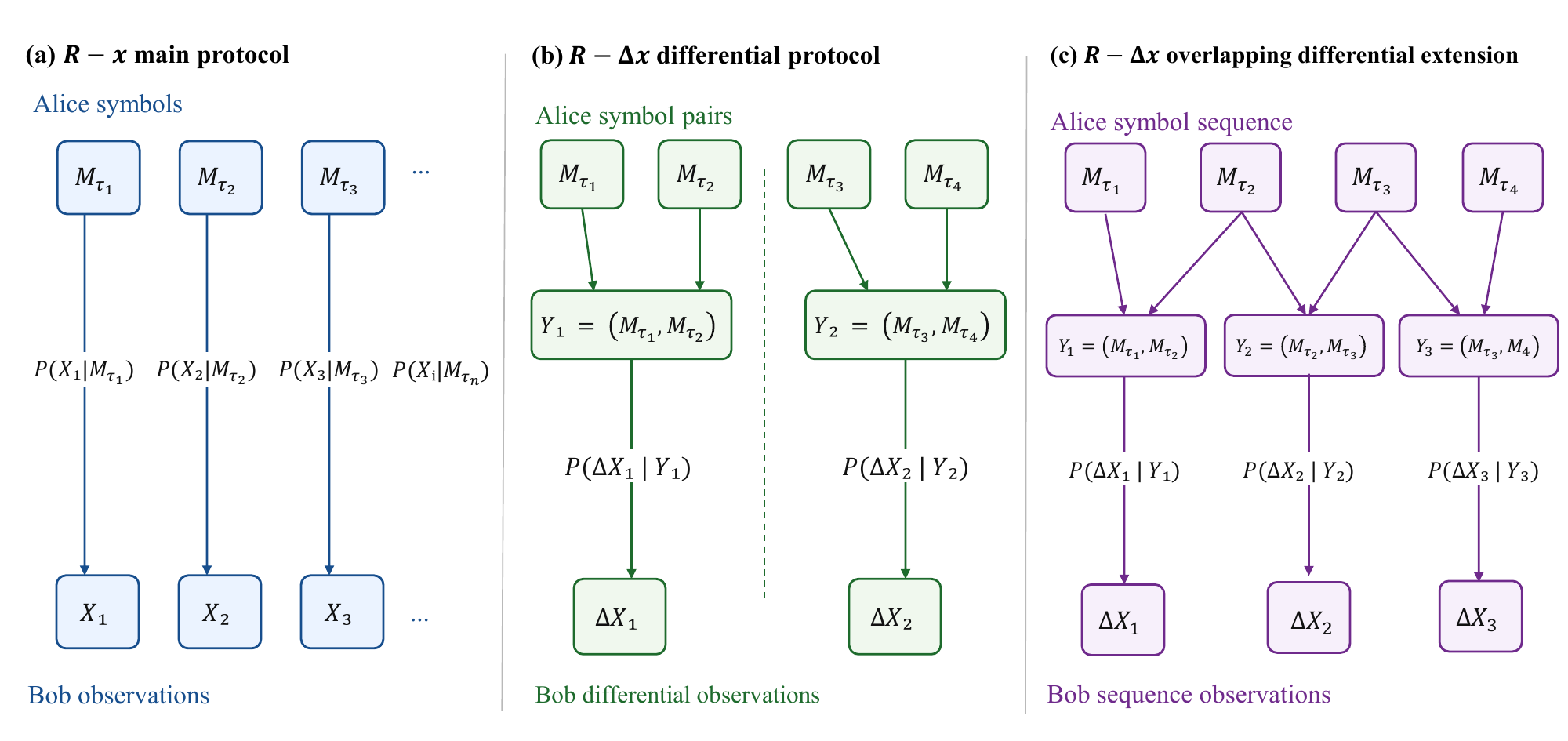}
\caption{
\textbf{Statistical structures of the main $R-x$ protocol and the $R-\Delta x$ differential protocol.}
(a) In the main $R-x$ protocol, each Alice symbol random variable $M_{\tau_i}$ is associated with Bob's detection-coordinate observation $X_i$ through the single-event conditional probability $P(X_i|{\tau_i})$, forming symbol-wise soft information. (b) In the non-overlapping $R-\Delta x$ differential protocol, adjacent accepted symbols form a symbol pair $Y_r=(M_{\tau_{2r-1}},M_{\tau_{2r}})$, which is associated with the differential observation $\Delta X_r$ through $P(\Delta X_r|Y_r)$, forming symbol-pair soft information. (c) In the overlapping differential extension, adjacent differential samples share Alice symbols and form a chain-coupled structure; post-processing must therefore be handled according to a sequence soft-input information model.
}
\label{fig:protocol-statistical-structure}
\end{figure}

The $R-\Delta x$ protocol is based on the differential observation variable defined in Section~\ref{subsec:differential-observation-variable}. Its core idea is to use the relative detection coordinate of adjacent accepted detection events instead of the absolute detection coordinate, thereby mitigating the influence of slowly varying center drift on the observation model. Unlike the single-symbol observation in the main $R-x$ protocol, a differential observation associates one Bob observation variable with a pair of Alice geometrical-configuration symbols. Figure~\ref{fig:protocol-statistical-structure} shows the statistical structures of the main $R-x$ protocol, the non-overlapping $R-\Delta x$ differential protocol, and the overlapping differential extension.

\subsection{Pairing of Accepted Detection Events}
\label{subsec:rdeltax-pairing}

After quantum transmission is completed, Bob publicly announces the time indices of accepted detection events. Let the emission time slots corresponding to accepted detection events be
\begin{equation}
\tau_1<\tau_2<\dots<\tau_n.
\end{equation}
Bob holds the detection-coordinate sequence
\begin{equation}
X_{1},X_{2},\dots,X_{n},
\end{equation}
and Alice holds the corresponding symbol sequence
\begin{equation}
M_{\tau_1},M_{\tau_2},\dots,M_{\tau_n}.
\end{equation}

We use non-overlapping pairing as the main version of the $R-\Delta x$ protocol. Specifically, let
\begin{equation}
\Delta X_r
=
X_{2r}-X_{2r-1},
\quad
r=1,2,\dots,\left\lfloor \frac{n}{2}\right\rfloor .
\end{equation}
The Alice symbol pair corresponding to the $r$-th differential sample is
\begin{equation}
Y_r
=
(M_{\tau_{2r-1}},M_{\tau_{2r}}).
\end{equation}

To ensure that the slowly varying drift is approximately the same within each event pair, the protocol sets a maximum pairing time interval $T_\Delta$. Only event pairs satisfying
\begin{equation}
t_{\tau_{2r}}-t_{\tau_{2r-1}}
\leq
T_\Delta
\end{equation}
enter the raw differential data. Event pairs that do not satisfy this condition are discarded or used for parameter estimation.

Non-overlapping pairing reduces sample utilization, but makes different differential samples closer to independent, which facilitates information reconciliation and security analysis. Overlapping differential construction can improve sample utilization, but introduces shared events and sample correlations; we treat it as an extension rather than as the main protocol version. The pairing-screening rule is part of the public protocol specification. If the effective pairing probability depends on the symbol pair $(a,b)$, this selection bias should enter the estimation of the symbol-pair prior $\tilde{P}(a,b)$ and be constrained separately in parameter estimation.

\subsection{Differential Conditional Probability}
\label{subsec:rdeltax-conditional-probability}

Given a symbol pair $(a,b)$, the corresponding source separations are $R_a$ and $R_b$. Under non-overlapping pairing and the conditional-independence approximation, the continuous conditional probability of the differential variable is
\begin{equation}
P(\Delta x|R_a,R_b,\phi_a,\phi_b)
=
\int
P_{\mathrm{obs}}(x|R_a,\phi_a)
P_{\mathrm{obs}}(x+\Delta x|R_b,\phi_b)
\,dx.
\end{equation}
Here $P_{\mathrm{obs}}(x|R_m,\phi)$ denotes the single-event observation distribution after center correction or marginalization has been completed.

In the discrete pixel model, if the two events fall into pixels $j$ and $k$, respectively, then
\begin{equation}
\Delta x_{kj}
=
x_k-x_j.
\end{equation}
The corresponding discrete differential probability is
\begin{equation}
P(\Delta X=\delta|R_a,R_b,\phi_a,\phi_b)
=
\sum_{\substack{j,k:\\x_k-x_j=\delta}}
p_j(R_a,\phi_a)
p_k(R_b,\phi_b).
\end{equation}
If the detector pixels are equally spaced, the pixel-index difference $d=k-j$ can be used to represent the differential result:
\begin{equation}
P(D=d|R_a,R_b,\phi_a,\phi_b)
=
\sum_j
p_j(R_a,\phi_a)
p_{j+d}(R_b,\phi_b),
\end{equation}
where out-of-range terms are taken to be zero. This expression shows that the differential distribution is obtained from the cross-correlation of the detection-coordinate distributions corresponding to two geometrical configurations.

If a residual drift difference $\delta s$ exists between the paired events, the differential conditional probability can be written as
\begin{equation}
P(\Delta x|R_a,R_b,\phi_a,\phi_b,\delta s)
=
\int
P_{\mathrm{obs}}(x|R_a,\phi_a)
P_{\mathrm{obs}}(x+\Delta x-\delta s|R_b,\phi_b)
\,dx.
\end{equation}
When $\delta s$ has only a statistical description, the marginalized model
\begin{equation}
P(\Delta x|R_a,R_b,\phi_a,\phi_b)
=
\int
P(\Delta x|R_a,R_b,\phi_a,\phi_b,\delta s)
\rho_\Delta(\delta s)\,d(\delta s).
\end{equation}
is used.

\subsection{Differential Soft-Input Information and Reconciliation}
\label{subsec:rdeltax-soft-information}

For the $r$-th non-overlapping differential sample, Bob observes
\begin{equation}
\Delta x_r
=
X_{2r}-X_{2r-1}.
\end{equation}
The corresponding Alice symbol pair is
\begin{equation}
Y_r
=
(M_{\tau_{2r-1}},M_{\tau_{2r}}).
\end{equation}
Bob computes the log-likelihood for all candidate symbol pairs $(a,b)$:
\begin{equation}
L_r(a,b)
=
\log
P(\Delta x_r|R_a,R_b,\phi_{\tau_{2r-1}},\phi_{\tau_{2r}}).
\end{equation}
Including the pair prior gives, then
\begin{equation}
\Gamma_r(a,b)
=
\log\pi_a+\log\pi_b
+
L_r(a,b).
\end{equation}

Under non-overlapping pairing, each differential sample corresponds to a $K^2$-ary pair symbol. Bob's soft information is
\begin{equation}
\boldsymbol{\Gamma}_r
=
\{\Gamma_r(a,b)\}_{a,b=0}^{K-1}.
\end{equation}
Alice holds the pair-symbol sequence
\begin{equation}
Y^{n_\Delta}
=
(Y_1,Y_2,\dots,Y_{n_\Delta}).
\end{equation}
Bob holds the corresponding soft-input information sequence
\begin{equation}
\boldsymbol{\Gamma}^{n_\Delta}
=
(\boldsymbol{\Gamma}_1,\boldsymbol{\Gamma}_2,\dots,\boldsymbol{\Gamma}_{n_\Delta}).
\end{equation}
The two parties then perform soft-input information reconciliation over the $K^2$-ary pair-symbol alphabet, so that Bob recovers Alice's symbol-pair sequence. After recovery, the symbol pairs can be expanded in a fixed order into Alice's geometrical-configuration symbol sequence.

\subsection{Overlapping Differential Extension}
\label{subsec:rdeltax-overlapping-extension}

Let $\tau_1,\tau_2,\dots,\tau_n$ denote the retained accepted detection time slots. If all adjacent accepted detection events are used to construct differential variables, then
\begin{equation}
\Delta X_\ell=
X_{\ell+1}-X_{\ell},
\quad
\ell=1,2,\dots,n-1.
\end{equation}
The corresponding observed value is denoted by $\Delta x_\ell$. In this construction, the $\ell$-th differential observation couples the adjacent symbol pair
\begin{equation}
\left(M_{\tau_\ell},M_{\tau_{\ell+1}}\right),
\end{equation}
and adjacent differential samples share the intermediate detection event. For example, both $\Delta X_\ell$ and $\Delta X_{\ell+1}$ depend on $X_{\ell+1}$. Therefore, overlapping differential variables generally cannot be treated as conditionally independent samples.

Under the calibrated single-event conditional probability model $p_X(x|M)$, and assuming that the accepted detection positions before differencing are conditionally independent given the symbol sequence, the exact posterior distribution of Alice's symbol sequence can be written as
\begin{equation}
P\left(
M_{\tau_1},\dots,M_{\tau_n}
\middle|
\Delta x_1,\dots,\Delta x_{n-1}
\right)=
\frac{1}{Z}
\mathcal{L}_{\mathrm{ov}}
\left(
\Delta x_1,\dots,\Delta x_{n-1}
\middle|
M_{\tau_1},\dots,M_{\tau_n}
\right)
\prod_{\ell=1}^{n}
\pi_{M_{\tau_\ell}},
\end{equation}
where $Z$ is the normalization factor, and the joint likelihood of the overlapping differential observations is
\begin{equation}
\mathcal{L}_{\mathrm{ov}}
\left(
\Delta x_1,\dots,\Delta x_{n-1}
\middle|
M_{\tau_1},\dots,M_{\tau_n}
\right)=
\int_{\mathcal{X}}
p_X\left(x\middle|M_{\tau_1}\right)
\prod_{\ell=1}^{n-1}
p_X\left(
x+\sum_{j=1}^{\ell}\Delta x_j
\middle|
M_{\tau_{\ell+1}}
\right)
,dx.
\end{equation}
For a discrete pixelated detector, the integral should be replaced by a summation over the initial pixel position.

If only pairwise edge likelihoods are retained, one may define
\begin{equation}
q_\ell
\left(
\Delta x_\ell
\middle|
M_{\tau_\ell},M_{\tau_{\ell+1}}
\right)=
\int_{\mathcal{X}}
p_X\left(x\middle|M_{\tau_\ell}\right)
p_X\left(x+\Delta x_\ell\middle|M_{\tau_{\ell+1}}\right)
,dx.
\end{equation}
The corresponding chain-factor approximation is then
\begin{equation}
\widetilde{P}\left(
M_{\tau_1},\dots,M_{\tau_n}
\middle|
\Delta x_1,\dots,\Delta x_{n-1}
\right)
\propto
\prod_{\ell=1}^{n-1}
q_\ell
\left(
\Delta x_\ell
\middle|
M_{\tau_\ell},M_{\tau_{\ell+1}}
\right)
\prod_{\ell=1}^{n}
\pi_{M_{\tau_\ell}}.
\end{equation}
This form corresponds to an approximate sequence-estimation problem on a chain factor graph, where $q_\ell$ should be interpreted as a pairwise edge potential or pseudo-likelihood factor rather than as a strictly independent observation likelihood. The overlapping differential construction can improve sample utilization, but it introduces memory effects and correlated noise; we therefore treat it as a future extension rather than as the default post-processing structure of the main protocol.

\subsection{Protocol Features}
\label{subsec:rdeltax-features}

The $R-\Delta x$ protocol uses paired events as basic samples and is more robust against slowly varying drift, but it must handle symbol-pair likelihoods, pairing loss, and differential noise. Its operating regimes are discussed in Section~\ref{subsec:rx-rdeltax-applicability}.

\section{Parameter Estimation, Candidate Key Rates, and Security Considerations}
\label{sec:parameter-estimation-security}

\subsection{Security Setting and Public Information}
\label{subsec:security-setting-public-information}

We adopt a prepare-and-measure QKD scenario. Alice prepares single-photon spatial superposition states modulated by the geometrical configuration of the light source, Bob performs spatially resolved single-photon detection in the far field or Fourier plane, and Eve can arbitrarily access the quantum channel. The classical channel between Alice and Bob is authenticated. Eve can eavesdrop on the classical communication content, but cannot tamper with the communication without being detected.

The protocol rules are all treated as public information. The geometrical-configuration alphabet $\mathcal{R}$, the symbol prior distribution $\pi_m$, the phase calibration rules, the conditional probability model, the accepted-event indices, the parameter-estimation samples, the public information disclosed during error correction, and the privacy-amplification rules can all be known to Eve. Protocol security is therefore based on the constraints imposed by parameter estimation on Eve's information and on the compression performed by privacy amplification, with the modulation rules and post-processing algorithms treated as public protocol components.

The candidate key-rate expressions in this section are formulated under a trusted-device and ideal single-photon-source model. Alice's state-preparation device, Bob's detector response, and the random-number generation process are regarded as calibrated and consistent with the protocol model. Detector side channels, source side channels, detection-efficiency attacks, and measurement-device-independent security correspond to stronger implementation-security models and require separate modeling in extended frameworks. For example, if the actual system uses a weak coherent source, the decoy-state method must be introduced to estimate the single-photon contribution, and the additional risks caused by multi-photon components must be handled in the key rate.

In a protocol run containing $N$ emission time slots, Alice holds the symbol random variable sequence
\begin{equation}
M^N=(M_1,M_2,\dots,M_N).
\end{equation}
For the main $R-x$ protocol, Bob's main observation variable is the detection-coordinate sequence $X^n$ on the accepted detection events. For the $R-\Delta x$ differential protocol, Bob's main observation variable is the differential sequence $\Delta X^{n_\Delta}$. 

\subsection{Parameter Estimation}
\label{subsec:parameter-estimation}

The role of parameter estimation is to characterize the channel state using experimental observables and to provide a reference for subsequently constraining Eve's information. The basic estimators considered in this paper include detection rate, background fraction, interference visibility, conditional-distribution residuals, and, in the differential protocol, pairing rate and differential-distribution residuals.

For geometrical configuration $R_m$, the accepted detection rate is defined as
\begin{equation}
Q_m
=
\frac{n_m}{N_m},
\end{equation}
where $N_m$ is the number of time slots in which Alice sends $R_m$, and $n_m$ is the number of these time slots that produce accepted detection events. If $Q_m$ exhibits anomalous dependence on $m$, this indicates that different geometrical configurations may have experienced nonuniform loss, detection bias, or selective intervention.

The far-field interference visibility can be estimated from sampled data or calibration data. For a given $R_m$, it can be written as
\begin{equation}
V_m
=
\frac{
P_{\max}^{(m)}-P_{\min}^{(m)}
}{
P_{\max}^{(m)}+P_{\min}^{(m)}
}.
\end{equation}
In practical processing, the effective visibility $\hat{V}_m$ can also be obtained by fitting the observed distribution $P_{\mathrm{obs}}(x|R_m,\phi,s)$. A decrease in visibility reduces Bob's distinguishability of geometrical-configuration symbols and may reflect coherence degradation, turbulence-induced perturbations, or perturbations introduced by eavesdropping.

The conditional-distribution residual is used to test the consistency between the observed distribution and the calibrated model. Let the empirical distribution obtained by sampling be $\hat{p}_j(R_m)$ and the calibrated distribution be $p_j(R_m,\phi)$. Define
\begin{equation}
\Delta_m
=
\sum_{j=1}^{J}
\left|
\hat{p}_j(R_m)-p_j(R_m,\phi)
\right|.
\end{equation}
If $\Delta_m$ exceeds the range allowed by statistical fluctuations, this indicates that the conditional probability model is inconsistent with the actual observations; the protocol run should then be aborted or more conservative security parameters should be adopted.

For the non-overlapping-pairing $R-\Delta x$ differential protocol, the fraction of accepted differential samples formed per emission time slot must also be estimated:
\begin{equation}
p_\Delta
=
\frac{n_\Delta}{N},
\end{equation}
where $n_\Delta$ is the number of accepted non-overlapping differential samples retained after time-interval screening. The differential-distribution residual can be defined as
\begin{equation}
\Delta_{ab}^{(\Delta)}
=
\sum_{\delta}
\left|
\hat{P}(\delta|R_a,R_b)
-
P(\delta|R_a,R_b)
\right|.
\end{equation}
This quantity is used to test whether the differential conditional probability model can still describe the actual observed data.

In addition to the key path, Bob may also retain a small number of samples for near-field or conjugate-space checks. If the protocol includes a conjugate-check path, Bob may randomly select a fraction of photons to enter a near-field or other conjugate-space measurement path. Let the empirical distribution deviations of the key path and the check path be $\Delta_p$ and $\Delta_x$, respectively. Then they can be used together with $Q_m$, $\hat{V}_m$, $\Delta_m$, and $\Delta_{ab}^{(\Delta)}$ as inputs to parameter estimation. The conjugate-check path provides additional perturbation diagnostic information and serves as an input to parameter estimation.

\subsection{Mutual Information and Reconciliation Efficiency}
\label{subsec:mutual-information-and-reconciliation}

An asymptotic mutual-information method is used to estimate the candidate key rate. The reconciliation efficiency is denoted by $\beta$, and the loss of the practical error-correction process relative to the ideal Shannon limit is included in the $\beta I$ term.

For the main $R-x$ protocol, key generation uses only accepted detection events. Let $D=1$ denote that a certain emission time slot produces an accepted detection event. If Alice's raw prior distribution is $\pi_m$ and the accepted detection rate of geometrical configuration $R_m$ is $Q_m$, then the symbol prior in the accepted-event set is
\begin{equation}
\tilde{\pi}_m
=
P(M=m|D=1)
=
\frac{\pi_m Q_m}{\sum_{n=0}^{K-1}\pi_n Q_n}.
\end{equation}
When the detection rate is approximately independent of the geometrical configuration, $\tilde{\pi}_m\simeq \pi_m$.

Let
\begin{equation}
\tilde{p}_j(R_m)
=
P(X=x_j|M=m,D=1)
\end{equation}
be the observation conditional probability conditioned on an accepted detection event. This conditional probability is determined by system calibration and parameter estimation, and includes experimental factors such as phase, center offset, background fraction, and detector response. Then the mutual information between Alice's symbol random variable $M$ and Bob's detection coordinate $X$ in a single accepted detection event is defined as
\begin{equation}
I_x
\equiv
I(M;X|D=1)
=
\sum_{m=0}^{K-1}
\sum_{j=1}^{J}
\tilde{\pi}_m \tilde{p}_j(R_m)
\log_2
\frac{
\tilde{p}_j(R_m)
}{
\sum_{n=0}^{K-1}
\tilde{\pi}_n \tilde{p}_j(R_n)
}.
\end{equation}
$I_x$ denotes the average amount of information provided by Bob's spatial detection-coordinate observation about Alice's geometrical-configuration symbol in each accepted detection event. When the conditional distributions corresponding to different $R_m$ are more easily distinguishable, $I_x$ is larger; conversely, when the conditional distributions overlap more strongly, $I_x$ is smaller.

For the non-overlapping-pairing $R-\Delta x$ differential protocol, the basic information unit is Alice's symbol pair and Bob's difference variable. Let
\begin{equation}
G=(M_1,M_2)
\end{equation}
denote Alice's symbol pair in one non-overlapping pair, and let $D_\Delta=1$ denote that this pair forms an accepted differential sample. The symbol-pair prior in the accepted differential samples is
\begin{equation}
\tilde{P}(a,b)
=
P(M_1=a,M_2=b|D_\Delta=1).
\end{equation}
The mutual information in each accepted differential sample is defined as
\begin{equation}
I_\Delta
\equiv
I(G;\Delta X|D_\Delta=1)
=
\sum_{a,b}
\sum_{\delta}
\tilde{P}(a,b)
\tilde{P}(\delta|R_a,R_b)
\log_2
\frac{
\tilde{P}(\delta|R_a,R_b)
}{
\sum_{c,d}
\tilde{P}(c,d)
\tilde{P}(\delta|R_c,R_d)
}.
\end{equation}
If $M_1$ and $M_2$ are independent at the transmitter and follow the prior distribution $\pi_m$, and if the screening process for accepted differential samples does not introduce a selection bias related to $(a,b)$, then
\begin{equation}
\tilde{P}(a,b)\simeq \pi_a\pi_b.
\end{equation}
If the screening process for accepted differential samples depends on the symbol pair, then $\tilde{P}(a,b)$ should be obtained by parameter estimation.

$I_\Delta$ is the mutual information defined per accepted differential sample. In this paper, the key rate is uniformly expressed per emission time slot. It should therefore be noted that the sample loss caused by non-overlapping pairing has already been included through the effective differential sample fraction $p_\Delta$ defined in parameter estimation.

The reconciliation efficiencies are denoted by $\beta_x$ and $\beta_\Delta$, respectively. For the main $R-x$ protocol, the Alice--Bob correlated information available for key extraction after information reconciliation is
\begin{equation}
\beta_x I_x.
\end{equation}
For the $R-\Delta x$ differential protocol, the corresponding effective correlated information is
\begin{equation}
\beta_\Delta I_\Delta.
\end{equation}

\subsection{Asymptotic Candidate Key Rates}
\label{subsec:candidate-key-rate}

The key rate is measured per emission time slot. For the main $R-x$ protocol, the average probability that an emission time slot produces an accepted detection event is
\begin{equation}
p_{\mathrm{det}}
=
P(D=1)
=
\sum_{m=0}^{K-1}\pi_m Q_m.
\end{equation}
The Alice--Bob mutual information in a single accepted detection event is $I_x=I(M;X|D=1)$. Therefore, the asymptotic candidate key rate of the main $R-x$ protocol is written as
\begin{equation}
r_x^{\mathrm{LB}}
=
p_{\mathrm{det}}
\left[
\beta_x I_x
-
\chi_x(M;E)
\right].
\end{equation}
Here $\beta_x$ is the reconciliation efficiency of the main $R-x$ protocol, and $\chi_x(M;E)$ is an upper bound on Eve's Holevo information about Alice's symbol random variable $M$. This Holevo information is measured per accepted detection event and is further constrained by the parameter-estimation results.

For the non-overlapping-pairing $R-\Delta x$ differential protocol, $p_\Delta=n_\Delta/N$ denotes the fraction of effective differential samples that can be formed per emission time slot. The Alice--Bob mutual information in a single accepted differential sample is $I_\Delta=I(G;\Delta X|D_\Delta=1)$, where $G=(M_1,M_2)$ is Alice's symbol pair in one non-overlapping pair. The corresponding asymptotic candidate key rate is written as
\begin{equation}
r_\Delta^{\mathrm{LB}}
=
p_\Delta
\left[
\beta_\Delta I_\Delta
-
\chi_\Delta(G;E)
\right].
\end{equation}
Here $\beta_\Delta$ is the reconciliation efficiency of the differential protocol, and $\chi_\Delta(G;E)$ is an upper bound on Eve's Holevo information about the symbol pair $G$.

Therefore, the task of parameter estimation is to provide an upper bound on $\chi_x(M;E)$ or $\chi_\Delta(G;E)$ from the detection rates, background fraction, visibility, conditional-distribution residuals, differential statistics, and conjugate-check results. If the corresponding right-hand side is not positive, the final secret key length for that round is set to zero.

A complete composable security analysis further needs to provide parameter-estimation confidence intervals, the reconciliation failure probability, the privacy-amplification failure probability, and finite-sample corrections to the upper bound on Eve's information. Equivalently, starting from finite-sample observation results, one needs to provide a computable lower bound on $H_{\min}^{\epsilon}(M_{\mathrm{raw}}|EC)$. These quantities constitute the key steps from the asymptotic candidate key rates to a final finite-key length formula.

\subsection{Bounding Eve's Information}
\label{subsec:eve-information-bound}

The main security consideration in GM-QKD is to quantitatively bound the information available to Eve about the geometrical-configuration symbols. Bob obtains information about $R_m$ through single-photon detection-coordinate statistics; Eve may also obtain related information by measuring, coupling to, or selectively intervening in the single-photon state in the quantum channel. Therefore, security must compare the Alice--Bob correlation with the information available to Eve.

Eve's influence may appear as changes in detection rate, geometrical-configuration-dependent loss, decreased interference visibility, anomalous conditional distributions, changes in the background distribution, increased differential noise, or distribution distortion in the conjugate-check path. The goal of parameter estimation is to transform these observations into a constraint set compatible with Eve's information. Formally, one can define the set of all Alice--Eve joint states compatible with the observed parameters as $\mathcal{S}_{\mathrm{obs}}$ and require
\begin{equation}
\chi(M;E)
\leq
\max_{\rho_{ME}\in\mathcal{S}_{\mathrm{obs}}}
\chi(M;E)_\rho .
\end{equation}
Here $\mathcal{S}_{\mathrm{obs}}$ is jointly constrained by the detection rates, visibility, conditional-distribution residuals, differential statistics, and conjugate-check results. All public classical information is denoted collectively by $C$. The final security analysis should be performed conditioned on $C$, namely, Eve's side information should be written as $EC$.

The Holevo constraint above defines a worst-case security optimization problem rather than a closed-form security bound. A complete security proof must derive an explicit computable upper bound on $\chi(M;E)$, or equivalently a lower bound on $H_{\min}^{\epsilon}(M_{\mathrm{raw}}|EC)$, from experimentally observable quantities. If this bound cannot be established with the required confidence, or if the resulting candidate key rate is non-positive, the protocol run is aborted and no final key is generated.

\section{Discussion}
\label{sec:discussion}

\subsection{Physical Role of GM-QKD}
\label{subsec:physical-positioning}

GM-QKD uses the geometrical configuration of a coherent emitting structure as a modulation degree of freedom for single-photon state preparation. In the minimal double-source implementation, Alice selects the source separation $R_m$, thereby preparing different single-photon spatial superposition states; Bob observes the conditional probability distribution determined by $R_m$ in the far field or Fourier plane. Its core feature is that the source separation determines the spatial frequency of the far-field interference distribution.

The raw random variable of GM-QKD is Alice's geometrical-configuration symbol; Bob's detection-coordinate or differential observation provides statistical information about this symbol. Bob generates soft information through calibrated conditional probabilities, and the key is extracted through information reconciliation and privacy amplification. The role of the geometrical configuration is to provide a modulation approach based on spatial degrees of freedom.

Compared with conventional spatial-mode QKD, GM-QKD has a different starting point. Conventional spatial-mode QKD usually preselects a set of spatial modes as the encoding basis; GM-QKD instead starts from the transmitter geometry, with geometrical parameters determining the actually prepared states and their probability distributions after propagation. The double-source separation $R$ used in this paper is the minimal implementation of this idea, while more complex multi-source arrays, two-dimensional aperture arrangements, and programmable emitting structures can be regarded as high-dimensional extensions of GM.

\subsection{\texorpdfstring{Operating Regimes of the $R-x$ and $R-\Delta x$ Protocols}{Operating Regimes of the R-x and R-Delta x Protocols}}
\label{subsec:rx-rdeltax-applicability}

The main $R-x$ protocol directly uses Bob's absolute detection coordinate as the observation variable. Its conditional probability model is clear, and its post-processing complexity is relatively low. It is suitable for scenarios in which the receiver is stable, the center offset can be estimated, and the calibrated model is reliable, and it is the basic protocol model of GM-QKD.

The $R-\Delta x$ differential protocol is intended for free-space links with significant slowly varying drift. It uses the difference variable between adjacent accepted detection events and can mitigate the influence of common-mode drift on absolute detection coordinates. Its cost is the introduction of symbol-pair coupling, pairing loss, enhanced differential noise, and sample correlations. Therefore, when beam tracking and center estimation are reliable, $R-x$ is more direct; when slowly varying drift becomes the main limitation, $R-\Delta x$ can provide a more robust observation method.

\subsection{Main Open Problems}
\label{subsec:open-problems}

The physical model of GM-QKD and the two candidate protocols, namely $R-x$ and $R-\Delta x$, define the scope of the present framework. Subsequent work centers on computable security bounds, finite-key length formulas, and performance-optimization methods oriented toward experimental parameters.

First, it is necessary to establish a computable relationship from observed statistics to an upper bound on Eve's information. In Section~\ref{subsec:eve-information-bound}, we formulate the security problem as a worst-case optimization over a constraint set, namely, maximizing Eve's Holevo information about Alice's symbols over all attacks compatible with the detection rates, visibility, conditional-distribution residuals, differential statistics, and conjugate-check results. Converting this formal statement into a concrete computable upper bound and combining it with the Alice--Bob mutual information gives a conservative estimate of the securely extractable key rate.

Second, the geometrical-configuration alphabet needs to be optimized. The conditional distributions corresponding to different $R_m$ must be sufficiently distinguishable for Bob to obtain useful mutual information from the detection coordinate or differential observation. At the same time, increasing Alice--Bob distinguishability may also increase Eve's distinguishability under the allowed attack model. The relevant optimization target is therefore not distinguishability alone, but the candidate key-rate balance between the reconciliation term and the upper bound on Eve's information.

Third, the post-processing methods of the $R-\Delta x$ differential protocol need to be systematically evaluated. Non-overlapping pairing gives a clear symbol-pair model,
\begin{equation}
G=(M_1,M_2),
\end{equation}
and its mutual information, information reconciliation, and upper bound on Eve's information can all be defined per accepted differential sample. Overlapping differential construction can improve sample utilization, but adjacent differential variables share the same detection event, forming a sequence observation problem with memory. Future work needs to use factor graphs, soft-input error correction, or decoding methods for channels with memory to compare non-overlapping pairing and overlapping differential construction in terms of achievable code rate, error-correction complexity, and complexity of security analysis.

Fourth, performance bounds under free-space channels need to be established. Loss, turbulence, and background events jointly change the effective detection rate, interference visibility, and conditional distributions. Turbulence not only causes center drift, but also produces phase perturbations, wavefront distortion, spot broadening, and envelope changes. The $R-\Delta x$ differential protocol can suppress slowly varying common-mode drift, but cannot replace parameter estimation for visibility degradation and distribution distortion. Therefore, future experimental evaluation should separately examine the influence of center drift, visibility degradation, background fraction, and conditional-distribution residuals on the Alice--Bob mutual information and the constraint on Eve's information.

Fifth, finite-key effects need to be handled after the asymptotic security analysis is completed. In the finite-key case, parameter estimation should use confidence intervals, information reconciliation should use the actual length of public information, and privacy amplification should provide explicit security parameters.

\subsection{Experimental Considerations}
\label{subsec:experimental-considerations}

Experimentally, Alice can implement the discrete selection of source separation $R_m$ using a tunable double-aperture structure, a spatial light modulator, an integrated optical-switch array, or a programmable light-source array. Bob can use SPADs or other detectors with single-photon sensitivity and spatial resolution to record far-field detection coordinates.

System calibration should include at least three parts: calibration from geometrical configuration to far-field conditional probability distribution, calibration of detector pixel response, and calibration of center offset and background-event statistics. For free-space links, clock synchronization, beam tracking, background-light suppression, and a stable phase reference are also required.

The choice of the geometrical-configuration alphabet must also match the fringe period, detector pixel size, and effective field of view. The conditional distributions corresponding to adjacent $R_m$ should produce resolvable spatial-frequency differences within the detector field of view. At the same time, the shortest fringe period $\Lambda_m=\lambda L/R_m$ should be sampled by multiple pixels to avoid spatial aliasing and visibility-estimation bias. The maximum source separation is also jointly constrained by the transmitter aperture, receiver aperture, coherence preservation, and beam-overlap conditions.

In the $R-\Delta x$ differential protocol, the maximum pairing time interval $T_\Delta$ is a key parameter. If $T_\Delta$ is too large, the drift between adjacent events is no longer approximately the same; if $T_\Delta$ is too small, the effective pairing rate decreases. Therefore, $T_\Delta$ should be determined jointly according to the channel coherence time, detection rate, and turbulence strength.

\section{Conclusion}
\label{sec:conclusion}

This paper has proposed a candidate framework for QKD based on geometrical-configuration modulation. Using the double-source single-photon spatial superposition state as the basic physical model, the framework takes the source separation $R$ of Alice's coherent double source as the modulation variable, so that Bob can obtain soft information about the geometrical-configuration symbol through spatially resolved single-photon detection in the far field or Fourier plane.

On this basis, we construct the main $R-x$ protocol and the $R-\Delta x$ differential protocol. The former directly uses single-photon detection coordinates for information reconciliation, while the latter uses differential observations to mitigate the influence of slowly varying common-mode drift in free-space links. The two protocols provide, respectively, the basic reconciliation model of GM-QKD and a differential extension oriented toward free-space perturbations.

We further give a unified formulation of parameter estimation, mutual-information estimation, and asymptotic candidate key rates. Protocol security depends on the constraints imposed by experimental observables on Eve's information and on subsequent privacy amplification. Deriving an explicit computable upper bound on Eve's information from observed parameters, together with finite-key analysis and free-space experimental validation, remains the main direction for future work.


\bibliographystyle{unsrtnat}
\bibliography{refs}


\end{document}